\documentclass[12pt,twoside]{article}
\usepackage{epsfig}
\usepackage{graphicx}
\usepackage{dcolumn}
\usepackage{bm}

\def\mrm{\mathrm}
\def\ra{\rightarrow}

\newcommand{\GeV}{\ensuremath{\mathrm{Ge\kern -0.1em V}}}
\newcommand{\MeV}{\ensuremath{\mathrm{Me\kern -0.1em V}}}

\newcommand{\pb}{\ensuremath{\mathrm{pb}^{-1}}}

\newcommand{\bs}{\ensuremath{B_s^0}}
\newcommand{\bd}{\ensuremath{B_d^0}}
\newcommand{\bu}{\ensuremath{B^+}}
\newcommand{\jp}{\ensuremath{J/\psi}}
\newcommand{\mm}{\ensuremath{\mu^{+}\mu^{-}}}

\newcommand{\bsmm}{\ensuremath{\bs\ra\mm}}
\newcommand{\bdmm}{\ensuremath{\bd\ra\mm}}
\newcommand{\jpmm}{\ensuremath{\jp\ra\mm}}
\newcommand{\bjk}{\ensuremath{\bu\ra\jp K^{+}}}
\newcommand{\bsjphi}{\ensuremath{\bs\ra\jp \phi}}

\newcommand{\brbsmm}{\ensuremath{\mathrm{BR}(\bsmm)}}
\newcommand{\brbdmm}{\ensuremath{\mathrm{BR}(\bdmm)}}

\begin{document}

\begin{figure}
\leftline{\hfill \today}
\end{figure}
\vspace*{0.5in}

\begin{figure}
\leftline{\includegraphics[scale=0.5]{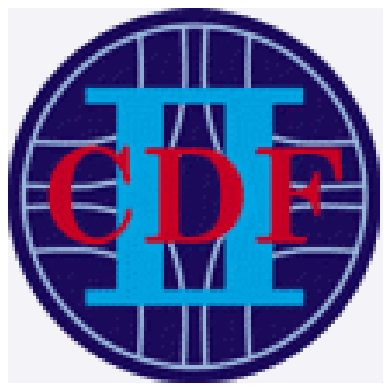}\hfill
}
\leftline{\includegraphics[scale=1.0]{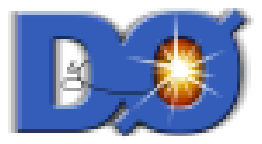}\hfill
}
\end{figure}

\begin{center}
  \boldmath
  {\LARGE\bf 
   A Combination of CDF and D\O~Limits on the Branching Ratio of
   $B_{s(d)}^0\rightarrow \mu^+ \mu^-$ Decays
   \vspace*{0.10in}\\[8pt]
  }
  \unboldmath
  \vfill
  {\large
     R.~Bernhard$^1$, D.~Glenzinski$^2$, M.~Herndon$^3$, T.~Kamon$^4$, 
     V.~Krutelyov$^4$, G.~Landsberg~$^5$, F.~Lehner$^1$, 
     C.-J.~Lin$^2$, S. Mrenna$^2$\\[15pt]
    {\small{
      {\em$^1$ University of Zurich}\\
      {\em$^2$ Fermi National Accelerator Laboratory}\\
      {\em$^3$ University of Wisconsin}\\
      {\em$^4$ Texas A\& M University}\\
      {\em$^5$ Brown University}
      }\\
\vspace*{0.3in}
    }
    for the CDF and D\O~Collaborations\\
  }
\end{center}
\vfill
\begin{abstract}
\noindent
  We combine the results of CDF and D\O~searches for the rare decays \bsmm\ and \bdmm .
  The experiments use $364\:\pb$ and  $300\:\pb$ of data respectively.
  The limits on the branching ratios are obtained by normalizing the estimated sensitivity to the decay
  \bjk\, taking into account the fragmentation ratios $f_u/f_{s(d)}$.
  The combined results exclude branching ratios of $\brbsmm > 1.5\times10^{-7}$ and
  $\brbdmm > 4.0\times10^{-8}$ at 95\% confidence level.  These are the most stringent limits on
  these decays at the present time.
\end{abstract}

\clearpage
\pagestyle{plain}

\section{Introduction}
\label{sec:intro}

  The CDF and D\O~experiments have previously reported on searches for the rare decay
\bsmm\ ~\cite{cdfprelim,d0prelim}.  CDF also directly searched for the decay \bdmm .
These decays are highly suppressed in the Standard Model of particle
physics with branching ratios of $\brbsmm = 3.5\times 10^{-9}$ and $\brbdmm = 1.0\times 10^{-10}$~\cite{smbr1}.  
However, decays of this type can be significantly enhanced 
in many scenarios beyond the Standard Model~\cite{uli1}.  
A combination of results leads to more stringent limits and is of considerable
interest in exploring the phase space of the models where strong enhancements for \bsmm\ or \bdmm\
are predicted.  In this note we report on a combination of limits in the
\bsmm\ and \bdmm\ decay channels.
\vspace*{0.10in}

  CDF and D\O~use similar methodologies to search for the \bsmm\ decay.
CDF applied the same methods to directly search for the decay \bdmm .
Each experiment looks for two oppositely charged muons in the $B_s^0$ and $B_d^0$ mass range
using dedicated triggers.  
CDF divides their dataset into two channels: the "Central" channel
consists of muon pairs reconstructed in the
pseudorapidity region, $|\eta| <0.6$, and the "Central-Extended" channel consists of dimuon 
events where one muon is reconstructed in the central region and the 
second muon in the extended muon system, $0.6 < |\eta| < 1.0$.
The two
channels have different sensitivities, therefore the optimization is
performed separately for each channel.
The branching ratio is computed by normalizing the number of signal 
events to the number of reconstructed \bjk~\cite{charge} events.
The branching ratio or limit is then calculated from the equation:
\begin{equation}\label{eq:intro}
\brbsmm = \frac{N^{obs}_{B_{s}}}{\alpha_{B_{s}}\epsilon^{\mrm{total}}_{B_{s}}}\cdot
          \frac{\alpha_{B^{+}}\epsilon^{\mrm{total}}_{B^{+}}}{N^{obs}_{B^{+}}}\cdot 
	  \frac{f_{u}}{f_{s}}\cdot
          {\rm BR}(\bjk)\cdot {\rm BR}(\jpmm),
\end{equation}
where:
\begin{itemize}
 \item $N^{obs}_{B_{s}}$ is the number of observed \bsmm\ candidates;
 \item $\alpha_{B_{s}}$ is the geometric and kinematic acceptance of the di-muon trigger for \bsmm\ decays;
 \item  $\epsilon^{\mrm{total}}_{B_{s}}$ is the total efficiency
(including trigger, reconstruction and analysis requirements) for \bsmm\ events in the acceptance;
  \item $N^{obs}_{B^{+}}$, $\alpha_{B^{+}}$, and
$\epsilon^{\mrm{total}}_{B^{+}}$ are similarly defined for \bjk\ decays;
 \item $f_{u}/f_{s}$ accounts for the different $b$-quark fragmentation 
probabilities and is:\\ $(0.397~\pm~0.010)/(0.107~\pm~0.011)~=~3.71~\pm~0.41$,
where the anti-correlation between the uncertainties has been accounted 
for~\cite{pdg};
  \item ${\rm BR}(\bjk)\cdot {\rm BR}(\jpmm) = 
(1.00\pm0.04)\times10^{-3}\:\cdot\:(5.88\pm0.10)\times10^{-2}
=(5.88\pm0.26)\times10^{-5}$ are used~\cite{pdg} .
\end{itemize}

The experiments normalize to the decay mode \bjk\ rather than to the $B_s$ decay \bsjphi . 
Normalizing to the decay \bjk\ is preferable since the
mode has higher statistics and the branching ratio and lifetime are well known
from the measurements at CLEO and the asymmetric $B$ factories.  In addition,
understanding the efficiency to detect \bsjphi\ events is complicated by
the presence of CP even and odd decay components which have different lifetimes. 
Finally normalizing to the mode \bsjphi\ does not eliminate the systematic uncertainty
from the ratio $f_{u}/f_{s}$ since current measurements of the branching ratio of \bsjphi\ are calculated using
the fragmentation ratio.
The expression for \bdmm\ is obtained by replacing \bs\ with
\bd\ and the fragmentation ratio with $f_{u}/f_{d}$ which is taken as unity.
The limits on the branching ratio \brbsmm\ are also calculated using the ratio
$f_{u}/f_{s}$~\cite{hfag} determined from Tevatron data based on the
CDF Run 1 analysis~\cite{cdffs}.  
\vspace*{0.10in}

To increase the sensitivity both experiments perform an optimization over the primary
discriminating variables, which are based on lifetime, compatibility between the momentum
vector of the candidate $B$ meson and the vector between the production and decay vertices, and
isolation in a cone around the candidate meson, where isolation is defined as the ratio
of sum of momenta of the candidate tracks divided by the sum of momenta the candidate tracks 
and other tracks in the cone.  
CDF optimizes for the best limit, and D\O~ for 
the best sensitivity, where sensitivity is defined as $\epsilon_{B_s}/(1+\sqrt{N_{BG}})$, and
$N_{BG}$ is the number of expected background events.
\vspace*{0.10in}

We use a Bayesian integration method to calculate the combined limits~\cite{stat}.
The method takes into account correlated and uncorrelated systematic uncertainties
between the two experiments, and between the two search channels of the CDF analysis.  
\vspace*{0.10in}

The key elements in calculating the limit are estimating central values and
uncertainties on the efficiency and acceptance
for the trigger and reconstruction, and estimating the background.
These elements are described in the next section, followed by a section on
the limit calculation and a final section where the results are summarized.


\section{Acceptance, Efficiency, Backgrounds and Associated Uncertainties}
\label{sec:eff}

Both experiments evaluate the acceptance and efficiency of their respective triggers,
reconstruction code and discriminating variables.  By normalizing the limit to the measured 
decay \bjk\ most of the systematic
uncertainties of these estimates cancel in the ratio.  
The sources of systematic uncertainties in the CDF and  D\O~analyses are discussed in
detail in references~\cite{cdfprelim,d0prelim}.
Systematic uncertainties on the estimates of the acceptance and efficiency 
are uncorrelated since they are
based on estimates of trigger and reconstruction efficiencies for different detectors.
Systematic uncertainties associated with the production of $B$ mesons are
possibly correlated since similar event generation programs are used in both analysis.  
These systematic uncertainties
affect the estimates of the efficiencies of the primary  discriminating variables.
However, the variables that are used
are qualitatively different.  For instance, the strongest discriminator in each analysis is
the lifetime variable. The CDF analysis uses the proper decay time in 3D
while D\O~selects events based on the 2D transverse decay length significance distribution.  
The discriminating variable used by D\O~is correlated with momentum where
the CDF variable is uncorrelated.
Therefore the uncertainties on $f_{u}/f_{s}$ and ${\rm BR}(\bjk)\cdot {\rm BR}(\jpmm)$ are
100\% correlated between the two experiments and all other uncertainties are treated as
uncorrelated. 

The experiments evaluate the background using sideband events.  The uncertainties on the backgrounds are
dominated by the statistical uncertainties.  The acceptance, efficiency, and background 
numbers are summarized in Table~\ref{tab:summary}.

\begin{table}[htb]
\begin{center}
{\footnotesize
\begin{tabular}{|c||cr|cr|cr|} \hline
            & \multicolumn{2}{c|}{CDF: Central}  
            & \multicolumn{2}{c|}{CDF: Central-Extended}  
            & \multicolumn{2}{c|}{D\O~} \\ \hline
 Luminosity
            & \multicolumn{2}{c|}{364\pb}  
            & \multicolumn{2}{c|}{336\pb}  
            & \multicolumn{2}{c|}{300\pb} \\ \hline\hline

 ($\frac{\alpha_{\bu} \cdot \epsilon_{\bu}^{total}}{\alpha_{\bs} \cdot \epsilon_{\bs}^{total}} $)
            & $0.852\pm0.084$ &($\pm9.9\%$) 
            & $0.485\pm0.048$ &($\pm~9.9\%$)
            & $0.247\pm0.019$ &($\pm~7.7\%$)  \\ \hline
 $N^{obs}_{\bu}$
            & $1785\pm 60$     &($\pm~3.4\%$) 
            & $696\pm 39$      &($\pm~5.6\%$) 
            &  $906\pm 41$     &($\pm~5.0\%$)  \\ \hline

 Uncor. Uncer.
            & \multicolumn{1}{c}{}
            & \multicolumn{1}{c|}{$(\pm 10.4\%)$}
            & \multicolumn{1}{c}{}
            & \multicolumn{1}{c|}{$(\pm 11.3\%)$}
            & \multicolumn{1}{c}{}
            & \multicolumn{1}{c|}{$(\pm ~9.2\%)$} \\ \hline

 $f_{u} / f_{s}$
            & \multicolumn{1}{c}{}
            & \multicolumn{2}{c}{$3.71\pm0.41$ } 
            & \multicolumn{1}{c}{($\pm 11.0\%$)} 
            & \multicolumn{2}{c|}{} \\ \hline

 { ${\rm BR}(\bjk$}
            &\multicolumn{6}{c|}{} \\
 { $\ra\mm K^{+})$ }
            & \multicolumn{1}{c}{}
            & \multicolumn{2}{c}{$(5.88\pm0.26) \times10^{-5}$ } 
            & \multicolumn{1}{c}{($\pm~4.0\%$)} 
            & \multicolumn{2}{c|}{} \\ \hline

 Cor. Uncer.
            & \multicolumn{1}{c}{}
            & \multicolumn{2}{c}{} 
            & \multicolumn{1}{c}{($\pm11.9\%$)} 
            & \multicolumn{2}{c|}{} \\ \hline

 $N_{back}^{expected}$
            & $0.81\pm0.12$     &($\pm14.8\%$) 
            & $0.66\pm0.13$     &($\pm19.7\%$) 
            & $4.3\pm1.2$      &($\pm27.9\%$)\\ \hline \hline

 $N^{obs}_{\bs}$
            & $0$     &
            & $0$     &
            & $4$     &  \\ \hline

 $ses$ $(\times10^{7})$
            & $1.04\pm0.16$   &($\pm15.8\%$) 
            & $1.52\pm0.25$   &($\pm16.4\%$)
            & $0.59\pm0.09$   &($\pm15.0\%$)   \\ \hline
 $ses$ $(\times10^{7})$
            & \multicolumn{2}{c}{$0.617~~~~~~~$ (CDF combined) } 
            & \multicolumn{2}{c|}{}
            &  &   \\ \hline \hline

 Exp Limit 90\% CL
            & \multicolumn{1}{c}{$3.5 \times10^{-7}$} &
            & \multicolumn{1}{c}{$5.6 \times10^{-7}$} &
            & \multicolumn{1}{c}{$3.5 \times10^{-7}$} &  \\ \hline

 Exp Limit 90\% CL
            & \multicolumn{2}{c}{$~~2.0 \times10^{-7}$ (CDF combined) }  
            & \multicolumn{2}{c|}{}
            & &  \\ \hline

\end{tabular}
}
\caption{\label{tab:summary} A summary of the inputs used in 
  equation~\ref{eq:intro} to estimate \brbsmm .  The relative 
  uncertainties are given parenthetically.  The single-event-sensitivity,
  $ses$, to a given branching ratio, corresponding to $N^{obs}_{\bs}=1$, 
  and the expected limit at 90\% Confidence Level (CL),
  under a hypothetical repetition of the experiments,
  are calculated using the inputs.
  The combined $ses$ and expected limit for the CDF ``Central'' and ``Central-Extended'' search channels are also given.}
\end{center}
\end{table}

\section{Limits}
\label{sec:Limits}
Using the Bayesian integration method we calculate the combined limits.
In the case of the D\O~search the dimuon mass signal region covers both the \bs\ and \bd .
The limit on the branching ratio that is extracted in one mode assumes that the
branching ratio in the other mode is zero, which results in a conservative limit.
This is the case in the framework of
Minimal Flavor Violating (MFV) SUSY models, where the
CKM matrix is the only source of flavor violation. In MFV SUSY the branching ratio for \bdmm\ is expected to be
suppressed relative to \bsmm\ by a factor of $|V_{td}/V_{ts}|^2$, making the contribution from
\bdmm\ negligible.
The 95\% confidence level (CL) limits on the branching ratio \brbsmm\ are reported in Table~\ref{tab:limitsbs}.
The combined limits are also calculated at 90\% CL.

\hspace*{-10.0cm}
\begin{table}[htb]
\begin{center}
\begin{tabular}{|c||cr|cr|cr|} \hline
            & \multicolumn{2}{c|}{CDF: Central}  
            & \multicolumn{2}{c|}{CDF: Central-Extended}  
            & \multicolumn{2}{c|}{D\O~} \\ \hline
 Luminosity
            & \multicolumn{2}{c|}{364~\pb}  
            & \multicolumn{2}{c|}{336~\pb}  
            & \multicolumn{2}{c|}{300~\pb} \\ \hline\hline

 $Limits$
            & \multicolumn{4}{c|}{}  
            & \multicolumn{2}{c|}{}    \\
  at 95\%CL
            & \multicolumn{4}{c|}{$\brbsmm\ < 2.0 \times 10^{-7}$}  
            & \multicolumn{2}{c|}{$\brbsmm\ < 3.9 \times 10^{-7}$}    \\ \hline

 $Combined$
            & \multicolumn{6}{c|}{$\brbsmm\ < 1.5 \times 10^{-7}$ at 95\% CL}    \\
 $Limits$
            & \multicolumn{6}{c|}{$\brbsmm\ < 1.2 \times 10^{-7}$ at 90\% CL}    \\ \hline

\end{tabular}
\caption{\label{tab:limitsbs} A summary of the limits on the branching ratio \brbsmm . }

\end{center}
\end{table}

For the branching ratio \brbdmm\ the factor $f_u/f_d$ is taken to be unity.
The efficiency for the $B_d^0$ channel is estimated to be 8\% lower than in the $B_s^0$
channel for the D\O~search, and the uncorrelated uncertainty is 10.2\%.
The CDF efficiency and errors are the same as for the \bsmm\ channel.
The limits on the branching ratio \brbdmm\ are reported in Table~\ref{tab:limitsbd}.

\begin{table}[htb]
\begin{center}
\begin{tabular}{|c||cr|cr|cr|} \hline
            & \multicolumn{2}{c|}{CDF: Central}  
            & \multicolumn{2}{c|}{CDF: Central-Extended}  
            & \multicolumn{2}{c|}{D\O~} \\ \hline
 Luminosity
            & \multicolumn{2}{c|}{364~\pb}  
            & \multicolumn{2}{c|}{336~\pb}  
            & \multicolumn{2}{c|}{300~\pb} \\ \hline\hline

 $Limits$
            & \multicolumn{4}{c|}{}  
            & \multicolumn{2}{c|}{}    \\
  at 95\%CL
            & \multicolumn{4}{c|}{$\brbdmm\ < 5.1 \times 10^{-8}$}  
            & \multicolumn{2}{c|}{$\brbdmm\ < 11.1 \times10^{-8}$}    \\ \hline

 $Combined$
            & \multicolumn{6}{c|}{$\brbdmm\ < 4.0 \times 10^{-8}$ at 95\% CL}    \\
 $Limits$
            & \multicolumn{6}{c|}{$\brbdmm\ < 3.2 \times 10^{-8}$ at 90\% CL}    \\ \hline

\end{tabular}
\caption{\label{tab:limitsbd} A summary of the limits on the branching ratio \brbdmm . }

\end{center}
\end{table}

The limits on the branching ratios \brbsmm\ and \brbdmm\ can be compared to
the values expected in the SM.  Currently the 95\% CL limit on \brbsmm\ is approximately
a factor of 50 larger than the SM branching ratio while the limit on
\brbdmm\ is approximately 400 times larger.

The uncertainty on the ratio $f_{u}/f_{s}$ is the largest systematic uncertainty in
limit calculation.
This error dominates the external sources of systematic uncertainties.
In order to facilitate recalculation of the limit as the measured value of $f_{u}/f_{s}$ is improved we give the
limits factoring out this contribution in Table~\ref{tab:limits3}.

\begin{table}[htb]
\begin{center}
\begin{tabular}{|c||cr|cr|cr|} \hline
            & \multicolumn{2}{c|}{CDF: Central}  
            & \multicolumn{2}{c|}{CDF: Central-Extended}  
            & \multicolumn{2}{c|}{D\O~} \\ \hline
 Luminosity
            & \multicolumn{2}{c|}{364~\pb}  
            & \multicolumn{2}{c|}{336~\pb}  
            & \multicolumn{2}{c|}{300~\pb} \\ \hline\hline

 $Limits$
            & \multicolumn{4}{c|}{}  
            & \multicolumn{2}{c|}{$~~~~\brbsmm/ (f_{u}/f_{s})~~~~$}    \\
  at 95\%CL
            & \multicolumn{4}{c|}{$\brbsmm/ (f_{u}/f_{s}) < 5.1 \times 10^{-8}$}  
            & \multicolumn{2}{c|}{$ < 10.1 \times10^{-8}$}    \\ \hline

 $Combined$
            & \multicolumn{6}{c|}{$\brbsmm/ (f_{u}/f_{s}) < 3.9 \times 10^{-8}$ at 95\% CL}    \\
 $Limits$
            & \multicolumn{6}{c|}{$\brbsmm/ (f_{u}/f_{s}) < 3.1 \times 10^{-8}$ at 90\% CL}    \\ \hline

\end{tabular}
\caption{\label{tab:limits3} A summary of the limits on the branching ratio $\brbsmm / (f_{u}/f_{s})$. }

\end{center}
\end{table}

The limit on branching ratio \brbsmm\ can also be calculated using an evaluation of the 
fragmentation fractions based only on Tevatron data.
This average, $f_{u}/f_{s} = (0.398\pm0.010)/(0.120\pm0.021) = 3.32\pm 0.59$,
differs slightly from the world average value and is dominated by
the comparison of the mixing probabilities between $B_s^0$ and $B_d^0$ mesons, $\chi_s$ and $\chi_d$.
Limits calculated using this value of $f_{u}/f_{s}$ 
are reported in Table~\ref{tab:limits2}.

\begin{table}[htb]
\begin{center}
\begin{tabular}{|c||cr|cr|cr|} \hline
            & \multicolumn{2}{c|}{CDF: Central}  
            & \multicolumn{2}{c|}{CDF: Central-Extended}  
            & \multicolumn{2}{c|}{D\O~} \\ \hline
 Luminosity
            & \multicolumn{2}{c|}{364~\pb}  
            & \multicolumn{2}{c|}{336~\pb}  
            & \multicolumn{2}{c|}{300~\pb} \\ \hline\hline

 $Limits$
            &\multicolumn{4}{c|}{ }
            &\multicolumn{2}{c|}{ }  \\ 
  at 95\%CL
            & \multicolumn{4}{c|}{$\brbsmm\ < 1.9 \times 10^{-7}$}  
            & \multicolumn{2}{c|}{$\brbsmm\ < 3.7 \times 10^{-7}$}    \\ \hline

 $Combined$
            &\multicolumn{6}{c|}{$\brbsmm\ < 1.4 \times 10^{-7}$ at 95\% CL}  \\
 $Limits$
            & \multicolumn{6}{c|}{$\brbsmm\ < 1.1 \times 10^{-7}$ at 90\% CL}    \\ \hline

\end{tabular}
\caption{\label{tab:limits2} A summary of the limits on the branching ratio \brbsmm\ using the Tevatron determination
of $f_{u}/f_{s}$. }

\end{center}
\end{table}

The limits were checked using independent programs for Bayesian limit integration
developed separately by the CDF and D\O~collaborations.  In addition, the cutoff of the assumed
prior distribution was varied and the distributions used for the uncertainties
in the calculation were taken as Gaussian distributions with a cutoff and Gamma functions.
The limits were found to vary by less than 5\% under these tests.

\section{Conclusion}
\label{sec:Conclusion}

We report on a combination of CDF and D\O~ limits on the branching 
ratios \brbsmm\ and \brbdmm .
The limits are obtained using a relative normalization to the decay
\bjk .
The individual limits are combined using a Bayesian integration technique that
takes into account correlated and uncorrelated systematic uncertainties
between the two experiments.  
The limits are found to be robust under
several tests.  The combined limits are:  
\vspace*{0.10in}

\begin{itemize}
  \item $\brbsmm < 1.5\times10^{-7}$ at 95\% CL
  \item $\brbsmm < 1.2\times10^{-7}$ at 90\% CL
  \item $\brbdmm < 4.0\times10^{-8}$ at 95\% CL
  \item $\brbdmm < 3.2\times10^{-8}$ at 90\% CL
\end{itemize}
\vspace*{0.10in}


\clearpage



\begin{thebibliography}{99.}
\bibitem{charge}
  Throughout this note, charge conjugation is implied.
\end{thebibliography}
\end{document}